\documentclass[a4paper]{JHEP3}
\usepackage{epsfig}
\usepackage{graphicx}
\usepackage{amssymb}
\usepackage{latexsym}

\newcommand{\be}{\begin{equation}}
\newcommand{\ee}{\end{equation}}
\newcommand{\bea}{\begin{eqnarray}}
\newcommand{\eea}{\end{eqnarray}}

\def\celsius{{\,^{\circ}{\rm C}}}
\newcommand{\tpd}[4]{{\left (\frac{\partial {#1}}{\partial {#2}}\right )^{#4}_{#3}}}
\usepackage{wrapfig}

\title{A critique of Sadi Carnot's work and a mathematical theory of the caloric.}
\author{N.D. Hari Dass
\\  Chennai Mathematical Institute, Chennai, India \\
CQIQC, Indian Institute of Science, Bangalore, India\\
Email: \email{dass@cmi.ac.in},\email{dass@cts.iisc.ernet.in}}

\abstract{In this work, Sadi Carnot's fundamental work is critically examined, and contrasted with modern thermodynamics. A mathematical
theory of his work is given on the basis of the observation of that in caloric theory dQ is a \emph{perfect differential.}
}


\begin{document}

\section{Sadi Carnot and the Motive Power of Heat}
\label{sec:sadicarnot}
Now we give a detailed account of Carnot's seminal work \emph{Reflexions sur la Puissance Motrice du Feu} published in 1824,
clear three decades ahead of the formulation of the first and second laws of thermodynamics. Fortunately, the english translation
\emph{Reflections on the Motive Power of Heat} is available \cite{carnot1824} making accessible to the english speaking world 
this great treasure of science, which, unfortunately, was ignored and antiquated even before its greatness was understood and 
appreciated. Its greatness was revealed to the world of science largely due to William Thomson(Lord Kelvin)'s epoch-making paper 
\emph{Account of Carnot's Theory} which appeared in 1849 \cite{kelvin1849}, nearly a quarter of century after Carnot's work was published. It is
remarkable that Thomson himself was a young man at the time, having just embarked on his scientific career.
The account given here is based both on the original work as well as Kelvin's paper.

Carnot's style of presentation would clearly be found cumbersome and confusing by the modern reader. It hardly has any equations,
and almost all the chief results, of which there really are very many, are derived in a verbose and descriptive manner. Lord
Kelvin's account is decidedly more modern both in its perspective, as well as in its presentation. It does make use of equations
as well as of calculus. It gives a \emph{mathematically precise} meaning to Carnot's axioms as well as his results. As a result of
this clarity, Kelvin is able to show that Carnot's theory contains even more remarkable results like what has come to be known
as the \emph{Clapeyron Equation}. But even
Kelvin's account may be found somewhat verbose. In this book, the author has given a \emph{succinct} mathematical theory
which covers all the principal conclusions of both Carnot and Kelvin. It also points out very clearly the experimental data
that would have been acid tests for the Caloric theory, an objective that was at the heart of Carnot's work.

Carnot makes the \emph{Caloric Theory} the cornerstone
of his analysis, and says about the former:\emph{'....This fact has never been called in question. It was first admitted without
reflection, and verified afterwards in many cases with the calorimeter. To deny it would be to overthrow the whole theory of
heat to which it serves as a basis. For the rest, we may say in passing, the main principles on which the theory of heat rests
require the most careful examination. Many experimental facts appear almost inexplicable in the present state of this theory'}.
Nevertheless, he expresses his disquiet
about this theory quite clearly in the course of his thesis. In fact, to quote him verbatim, \emph{'The fundamental law that we 
propose to confirm seems to us to require, however, in order to be placed beyond doubt, new verifications. It is based upon the
theory of heat as it is understood today, and it should be said that this foundation does not appear to be of unquestionable
solidity. New experiments alone can decide the question}. 

The student of modern science may then wonder the usefullness or the need for going into details of a work based on what is now
known to be incorrect, namely, the \emph{caloric} theory. The answer is that even such a student would be amazed to find how many
deep truths Carnot uncovered, based on wrong premises, that have nevertheless survived the later developments. It is indeed a
valuable lesson on how scientific theories are to be assessed. If one had concentrated only on these highly non-trivial aspects,
one may well have come to the conclusion, even to this date, that caloric theory may after all be right! 

The other important lesson that such a student
would learn from Carnot's work is the precision with which scientific questions can be formulated, and the objective way in
which they can be answered. He introduced techniques of scientific enquiry which were very original then, and are novel even now!
His focus was not so much on any actually practicable engine; rather, it was on narrowing in on the essentials of an ideal engine,
conceivable in the simplest way, unencumbred by needless complications. It was a precursor \emph{par execellence} to the later day
\emph{gedanken experiments}. In its simplicity and range of applicability, its closest intellectual equivalent is the \emph{Turing Machine}
of \emph{Computer Science}. Finally, Carnot's work is a testimony to the true spirit of enquiry, honestly raising doubts about
one's own work and demonstrating unswerving faith in experiments as the only arbiters of scientific truth. In fact, the author
believes that ones grasp of thermodynamics in particular, and science in general, will be significantly enriched through an 
understanding of Carnot's work.

Before proceeding to a description of his work, it is worth making note of the milestones in the subject that were already known
at the time of Carnot. The gas laws of Boyle-Mariotte, Charles-Gay Lussac and Dalton were firmly established. Specific heat
measurements by Clement and Desormes, as well as by Delaroche and Berard were used by him as important experimental inputs in
his analysis. The fact that \emph{sudden} compression of gases heats them up and equally, sudden rarefaction cools them
was known to him, and quantitative details provided by Poisson were used in his analysis. In modern terminology, this refers
to the so called \emph{adiabatic} processes. Carnot was well aware of Laplaces work on the speed of sound, which had, in a crucial
way, corrected the earlier calculations of Newton by correctly incorporating the adiabatic changes \cite{laplace:1816}.

\section{Carnot's objectives} 
His main objective was to investigate the \emph{motive power} of heat. In modern usage, this means the
ability of heat to provide mechanical work. The first important step in this direction was his recognition that the effects of
heat can be manifold like generation of electrical currents, chemical reactions, volume changes etc., and that to lay the foundations
of a particular effect of heat, it is necessary to \emph{imagine} phenomena where all other effects are absent. This is so that
the relation between cause(in this case heat), and the effect(in this case mechanical work), may be arrived at through certain
\emph{simple} operations.

Therefore he focusses on systems where the sole effect of heat is in producing mechanical work. In particular, where the mechanical
effects arise out of increases and decreases in volumes under varying conditions of temperatures and pressures. The two precise questions 
Carnot sets out to answer are:

(i)\emph{What is the precise nature of the thermal agency which produces mechanical work and nothing else?}

(ii)\emph{What is the amount of thermal agency needed to produce a given amount of work?}

With respect to the second question, he further raises the issue of whether there is any limit to the amount of work produced by
a given amount of the \emph{thermal agency}.

\section{Cycles} 
Carnot argued that as thermal agency not only produces work, but also alters the \emph{state} of the system, it is
in general not possible to \emph{disentangle} the two aspects of heat from each other. For example, when we heat a gas at constant
temperature, say, the gas expands leading to a change of state (to a new density) and at the same time work is performed by
the expanding gas against the pressure. To circumvent this, Carnot envisages a \emph{sequence of operations} that brings the body back
to its original state. That way, the body having been returned to its original state, the work performed can be related \emph{solely}
to the thermal agency. Thus he introduced the novel notion of \emph{cycles}. {\bf It is very important to emphasize that in the caloric theory,
the total heat absorbed or given out in a cycle has to be exactly zero}. Therefore whatever Carnot calls the \emph{thermal agency}, it
can not be the total heat absorbed.

Equivalently, \emph{heat is also a function of the state only} and ought to be representable as a singlevalued function of the state
$Q(V,T)$, $Q(T,P)$ etc. In particular $dQ$ is a \emph{perfect differential} and partial derivatives like $\tpd{Q}{V}{T}{}$ are
perfectly meaningful mathematically. This will be in great contrast to the situation in post-Carnot development of the subject,
which we shall name \emph{the new thermodynamics}, for ease of reference.

\section{Thermal agency}
Since in a cycle, the body returns to its original state, and as per the caloric theory
the amount of heat in a body depends only on its state, it follows that the total heat absorbed must necessarily be zero. What, then,
is the \emph{thermal agency} responsible for producing work at the end of a cycle, since it can not obviously be the heat absorbed?

Carnot observes, after a careful examination of various heat engines that perform work, that in all of them heat enters the engine
at a higher temperature, and leaves at a lower temperature. So he asserts that it is this \emph{fall} of the caloric from a higher
temperature to a lower temperature that characterizes the thermal agency. 
Hence, according to Carnot, work arises \emph{'not due to an actual consumption of caloric, but
to its transportation from a warm body to a colder body'}.
He then likens the situation to the manner in which a
\emph{water wheel} performs work. There the agency responsible for work is the water falling from a height; the work performed depends 
both on the \emph{quantity of water} falling, as well as \emph{the height} through which it falls. After the work has been performed,
the amount of water is \emph{unchanged}. 
\begin{figure}[htbp]
\centering
\includegraphics[width=2.5in]{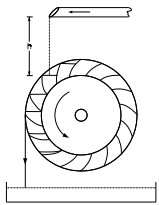}
\caption{The Water Wheel}
\label{fig:waterwheel}
\end{figure}

In fact, Carnot, in the light of the caloric theory, sees a perfect parallel between the water wheel and heat engines; the quantity of
water of the former corresponding to heat or the 'quantity of caloric' of the latter, the height of fall of the former corresponding
to the difference in the temperatures at which heat enters and leaves the engine. The caloric theory says that the amount of caloric, which
is neither creatable nor destructible, is invariable, and in the water wheel the amount of water is likewise. The comparison continues
to be apt even when we consider another subtle concept in Carnots work i.e reversibility, as we shall see soon.

\section{Ideal Heat Engines and Reversibility} 
The next important question raised by Carnot concerned the notion of \emph{the most
efficient} utilisation of the thermal agency in providing work. As is intutively obvious, there should be no \emph{wastages} of
the thermal agency. The following ingeneous criterion was found by Carnot: \emph{the most efficient(perfect) engine is such that,
whatever amount of mechanical effect it can derive from a certain thermal agency, if an equal amount be spent in working it backwards,
an equal reverse thermal effect will be produced}. This laid the foundation for the all important notion of {\bf reversibility} in
thermodynamics, and for that matter, quite generally in physics.
Recall our earlier characterization of a reversible change to be such that at the end of the
combined operation of the original process and its exact reverese, no changes should have occurred in the surroundings. 
Clearly, Carnot's criterion ensures this.

\begin{figure}[htbp]
\centering
\includegraphics[width=3.5in]{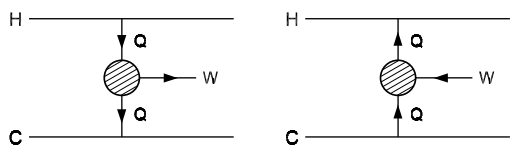}
\caption{A heat engine and its reverse in Carnot theory.}
\label{fig:carnotold}
\end{figure}
This criterion for reversibility can in principle allow irreversible changes of a type where \emph{less work} done in
reverese could restore the original thermal agency. It would be irreversible by the earlier criterion that the original
operation combined with the reverese would supply work to the surroundings at no cost of thermal agency. But such an
irreversible process can not be allowed as it amounts to a \emph{perpetual machine} which can supply \emph{indefinite}
amount of work at \emph{no cost}. Hence the irreversibility must be such that, when run in reverse, it must take
\emph{more} work to restore the original thermal agency. However, it is clear that perfectly reversible engines
permit the construction of \emph{perpetual machines}; but they can not perform any useful work. In practice, perfect
reversibility is any way not possible to achieve, and even perpetual machines of this limited kind are not possible.

Quite obviously, the reverse process should first of all be a \emph{physically realizable} process. Taking the water wheel as the
example, clearly the reverse process i.e of pumping water from a lower level to a higher level is certainly physically realizable.
Now if the wheel mechanism and other mechanisms involved in the water wheel are such that no work is dissipated in them, clearly
the reversibility criterion of Carnot will be fulfilled. In the case of the steam engine, wasteful effects like conduction of heat
through the walls of the boiler, for example, will degrade the efficiency for obtaining maximum possible work and therefore reversibility
requires their absence.

Therefore, the first important criterion for a \emph{perfect heat engine}
according to Carnot is that it should be \emph{reversible}. The criterion for reversibility enunciated by him is conceptually the 
\emph{simplest} and \emph{most straightforward}, with no hidden assumptions. For future reference, it is worth emphasizing that it 
is logically independent of the Second Law.

\section{Universality of Ideal Heat Engines}
Just using the notion of reversibility, and that of \emph{an ideal heat engine}, Carnot
proved a far reaching result concerning the universality of all ideal heat engines. It is indeed a stroke of a genius. The important
question posed by Carnot was whether the maximum efficiency of ideal heat engines depended on their design or not. In other words,
given \emph{ideal} heat engines of many kinds, will some of them be more efficient than others or not?

\begin{figure}[htbp]
\begin{minipage}[b]{0.5\linewidth}
\centering
\includegraphics[width=\linewidth]{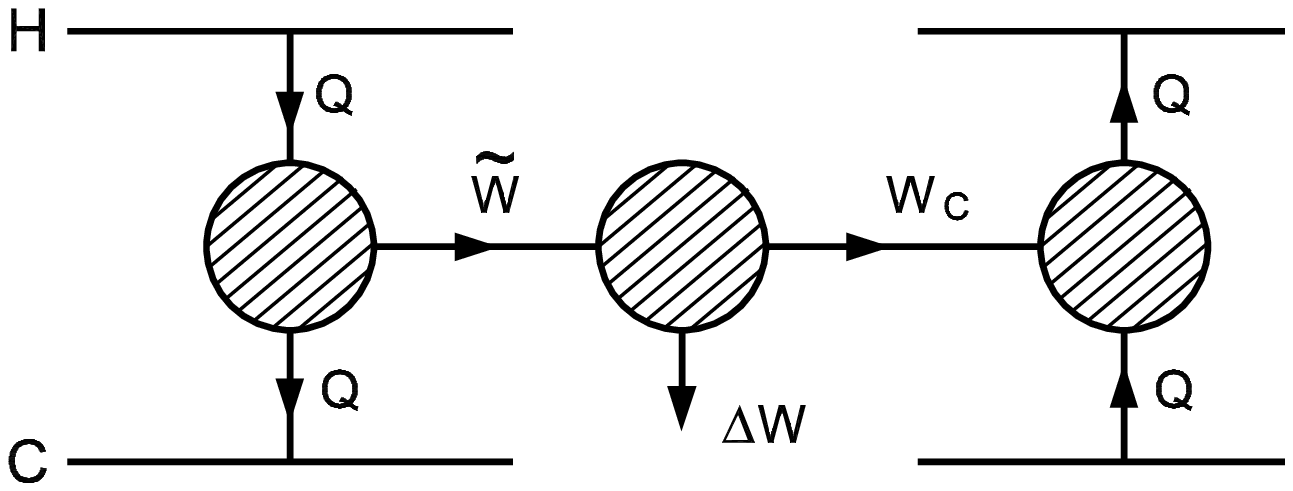}
\caption{Universality of carnot engines}
\label{fig:oldcarnotuniversality}
\end{minipage}
\begin{minipage}[b]{0.5\linewidth}
\centering
\includegraphics[width=0.8\linewidth]{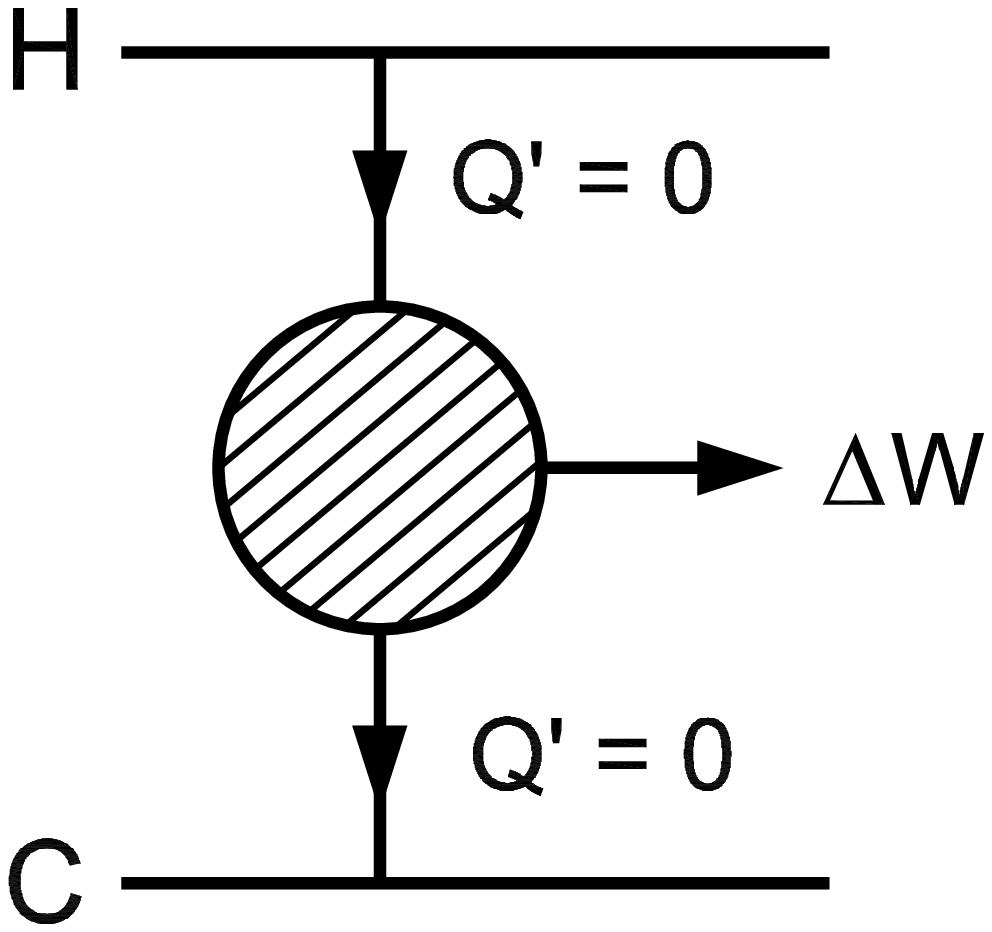}
\caption{Universality of carnot engines}
\label{fig:oldcarnotuniversality2}
\end{minipage}
\end{figure}
It would appear at first sight that the answer to such a very general question will not be easy to find, but Carnot solves
it in a truly ingeneous manner. Suppose there are two ideal heat engines $C, C^\prime$ such that for the same thermal
agency i.e a certain amount of heat Q falling through the temperatures $T_H, T_L$ with $T_H > T_L$, they deliver different
amounts of work $W, W^\prime$ with, say, $W^\prime > W$. Carnot considers splitting $W^\prime$ into $W+\Delta W$, and use
W to work C \emph{backwards}. Then, since C is ideal and hence reversible, run in reverse it will produce the same
thermal agency as C but in reverse i.e it will extract Q from $T_L$ and deliver all of it to $T_H$. The net effect of 
running $C^\prime$ and the reverese of C together is then that no \emph{net thermal agency} is used, yet there is net work 
$\Delta W$ produced. The cycles can be repeated \emph{forever} producing work \emph{indefinitely} out of \emph{nothing}.
This, Carnot argues, is \emph{inadmissible} and will violate the very basis of physics.

Consequently, Carnot arrives at what is perhaps one of the most remarkable scientific truths, namely, that \emph{all
ideal heat engines must deliver the same amount of work for a given amount of the thermal agency}. It would still be
possible to construct perpetual machines but of the kind that perform no useful work.

The true import of this \emph{universality} of all ideal heat engines is truly mind-boggling. For any given ideal
heat engine, this \emph{efficiency} i.e the amount of work performed for a given amount of thermal agency(it should
be carefully noted that \emph{efficiency} has a different meaning in the new thermodynamics), will
naturally depend on a number of properties of the substance employed in the engine. For example, in the case of steam
engines, it would involve such details as the \emph{latent heat}, {density} of both the liquid and vapor etc. Yet, the
combined dependence has to be such as to yield an universal efficiency. It has the further deep implication that,
knowing the value of this universal efficiency for one substance, say, air, would allow determination of some
property of another substance, say the latent of steam at some particular temperature, \emph{without} the need
for any experimental effort!

The only parallel one can think of is \emph{Einstein's Principle of Equivalence} in the theory of Gravitation; there
too, a \emph{theoretical principle}, if true, would determine the behaviour of all systems under the influence of 
gravitation if one knew their behaviour in accelerated frames. In that sense, Carnot's universality is also a
principle of equivalence i.e the equivalence of all ideal heat engines. One may even say that it is conceptually
on a firmer footing as its invalidation would lead to extraction of indefinite amount of work at no cost, and
hence the end of all physics, whereas Einstein's equivalence principle could in principle have been found to be
invalid experimentally!

\section{The Carnot Cycle} 
The cycle of reversible changes that Carnot envisaged as a means of addressing the question
of efficiency of ideal heat engines consists principally of \emph{four} stages in the following order: (i) an 
\emph{isothermal dilation} at a temperature $T_H$,
(ii) an \emph{adiabatic dilation} leading to a cooling from $T_H$ to $T_L < T_H$, (iii) an \emph{isothermal compression}
at $T_L$, and finally, (iv) an \emph{adiabatic
compression}. At the end of the fourth stage, the system is to return to its original thermodynamic state at the
beginning of (i).

\begin{figure}[htbp]
\centering
\includegraphics[width=1.5in]{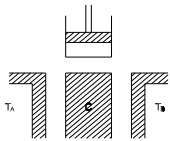}
\caption{Schematic setup of a Carnot cycle}
\label{fig:carnotsetup}
\end{figure}

There is however, a certain difficulty of an \emph{operational} nature as the Caloric theory requires that the \emph{heat
absorbed} during the first stage must \emph{exactly} match the \emph{heat relinquished} during the third stage. In other
words, while the end points B and C can be \emph{freely} chosen, the end point D has to be so chosen that the
last stage from D restores the system to its original starting point, and it is not clear how to identify such a D. To
circumvent this, Kelvin, and Maxwell, suggested variants of the cycle, which we shall take up shortly. 

Though
Carnot discussed the cycles both for an \emph{air engine} in which the working substance is any ideal gas, as well
as the steam engine where the working substance at every stage is water and steam in equilibrium, let us discuss
the cycle for the air engine first. This is because Carnot makes confusing statements about the realization of
a \emph{reversible} steam engine, even in an ideal sense. As Kelvin remarks in his commentary(he thanks Clapeyron
for the clarification), there are no such
difficulties and even for the steam engine, the same sequence of steps can be followed. The only thing to be kept
in mind is that at all stages the temperature of the water equals the temperature of vapor.

\begin{figure}
\centering
\includegraphics[width=2.5in]{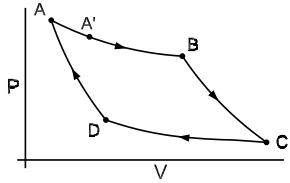}
\caption{Construction of Carnot Cycle.}
\label{fig:carnotconst}
\end{figure}
Carnot overcomes the operational difficulties(i.e of ensuring that the heat absorbed during (i) exactly matches
the heat relinquished during (iii)) as follows(see next figure): start with the system at $A^\prime$($P^\prime,V^\prime,
T_H$ and let it, under isothermal expansion, go to B$(P_B,V_B,T_H)$; then let (ii) be the adiabatic process taking
B to \emph{any} C($(P_C,V_C,T_L)$) such that C is at temperature $T_L$; in the next step, let (iii) isothermally 
take C to \emph{any} state D($(P_D,V_D,T_L)$); and let the adiabatic process (iv) take D to A($(P_A,V_A,T_H)$)
which is \emph{at the same temperature} as the starting temperature $T_H$. The operational difficulty now
manifests itself in that A need not necessarily be the same state as $A^\prime$, though both of them are at the
same $T_H$. But the point of Carnot is that an isothermal dilation starting from A \emph{has to} reach $A^\prime$,
and from then on simply retrace the earlier path $A^\prime B$. Now to get the Carnot cycle as prescribed earlier,
all one has to do is identify the entire path $AA^\prime B$ with the stage (i). Since no heat enters or leaves the system
through the phases (ii) and (iv), it follows that the heat absorbed during (i) has to necessarily match the heat given out
during (iii).

Carnot had also explicitly characterized stage (iii) to be such that it gives out all the heat the system had
absorbed during (i). Kelvin points out that spellt that way, this is the only part of the specification of the cycle
that is explictly sensitive to the correctness of the Caloric theory. Kelvin sought to free the description from this
by requiring the end point D of stage (iii) to be such that the fourth stage takes it to the starting state of (i).
Nevertheless, this does not solve the operational problem of locating such an end point. 
Maxwell's prescription,
which is completely operational, was to start from \emph{some} B at $T_H$, take it to \emph{some} C at $T_L$ via
an adiabatic dilation, take the system from C to  \emph{any} D, also at $T_L$, through an isothermal dilation, take 
D to \emph{some} A, as long as it is at $T_H$, and finally an isothermal dilation from A \emph{has to} take it to B.
It should be noted that this is pretty much the same strategy that Carnot also advocates.

The cycle for an ideal steam engine can also follow the same four stages with the important difference from gas
engines being in the fact that isothermal trajectories are also \emph{isobaric} i.e at constant pressure. This is
because the vapor pressure of \emph{saturated vapor} depends only on the temperature. This fact, as beautifully
analyzed by Clapeyron \cite{clapeyron:1834}
, actually allows the universal Carnot efficiency to be evaluated entirely in terms of physically
observable properties of the water-steam system, as will be discussed shortly.

The cycles are shown for the steam engine, as well as the gas engine in the next figure. For both of them, ABC is
the expansion phase and CDA, the contraction phase. A part of both of these is isothermal(AB,CD), and the other
adiabatic(BC,DA). During the isothermal phases, for a given volume, the pressure during the expansion(say, at $P_2$)
is always higher than the pressure during contraction(at $Q_2$). No such easy comparison is available during the
adiabatic phases. By drawing the verticals $P_1Q_1, P_3Q_3$ it is seen that at a given volume, the pressure during
the expansion is \emph{always} higher than the pressure during contraction, as shown in the next figure. 

\begin{figure}[htbp]
\centering
\includegraphics[width=3.5in]{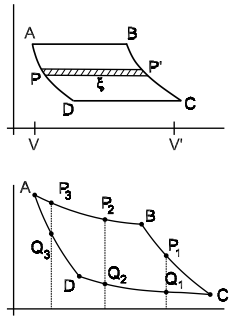}
\caption{Carnot cycles for air and steam engines.}
\label{fig:carnotoldcalc}
\end{figure}
Hence the mechanical work done \emph{by the system} during expansion is greater than the work done \emph{on the system} 
during contraction. This way, Carnot
concludes that \emph{net work} is done by the system at the end of the cycle. Kelvin
uses the graphical method to
show that the work done is the same as the \emph{area} of the curve ABCDA. The graphical methods 
are originally due to Clapeyron. For the modern student, that the area in the PV-diagram represents the
work is a straightforward consequence of calculus, but in the beginnings even this was a novel way of looking at
things.

\section{Carnot's style of analysis} 
As already mentioned, Carnot hardly made use of equations in his analysis, which
were mostly verbal augmented at most by simple arithmetical manipulations. Nevertheless, he made so many far
reaching conclusions with remarkable precision. We shall demonstrate this by a finer mathematical analysis whose
conclusions, as shown, coincided with his assertions made verbatim. But, to the modern reader, following Carnot's logic,
though impeccable, would indeed be tiresome. We illustrate this by his analysis of the relationship between
\emph{specific heats of ideal gases}, as an example.

First of all, he uses the Gay-Lussac law to argue that when a given mass of a gas is heated \emph{at constant pressure}
from $0\,^{\circ}{\rm C}$ to
1 $\celsius$, the \emph{fractional increase} in its volume is the \emph{same} for all gases and equals the fraction
$\frac{1}{267}$(the modern value would be closer to $\frac{1}{273}$). Therefore, the gas initially at $(P,V,0 \celsius)$
would go to $(P,V+\frac{V}{267},1\celsius$; the difference in heat between these two states is by definition the
\emph{specific heat at constant pressure}(for the given mass). 
He also uses the \emph{experimental data} of Poisson
that under an \emph{adiabatic compression} which raises the temperature of air by $1 \celsius$, its volume decreases
by a factor of $\frac{1}{116}$. Therefore, the heat content of the gas at $(P,V,0 \celsius)$ and at $(P^\prime, V-\frac{V}{116},1 \celsius)$
are the same(here $P^\prime$ is the pressure the gas would have at $1 \celsius$ when its volume is $V-\frac{V}{116}$).
On the other hand, if the gas had been heated at \emph{constant volume}, the heat required to raise the temperature
by $1 \celsius$ is, by definition, \emph{the specific heat at constant volume}(again for the given mass). Hence,
the specific heat at constant volume is \emph{also} the difference in heat between the states $(P^{\prime\prime},V,1 \celsius)$
and $(P^\prime, V-\frac{V}{116},1 \celsius)$. Now these two states are at the \emph{same} temperature but at \emph{different
volumes}. Carnot observes that the difference in their heat must be \emph{proportional} to the difference in thie volume
$\frac{V}{116}$. On the other hand, by similar reasoning, the specific heat at constant pressure will equal the difference
in heat between the states $(P^\prime,V-\frac{V}{116},1 \celsius)$ and $(P,V+\frac{V}{267},1 \celsius)$; these are
also at the same temperature, and therefore, the difference in their heat must also be proportional to the difference
in their volume, which is now $\frac{V}{116}+\frac{V}{267}$. It should be emphasized that the proportionality factor
is the \emph{same} as before. Let us call it X, for ease of reference.

From this rather verbose analysis, he rightly concludes that the ratio of the specific heat at constant pressure to the
specific heat at constant volume is $1 +\frac{116}{267}$ i.e the constant pressure specific heat is always greater
than the constant volume specific heat. This is usually attributed to \emph{First Law}, but Carnot's analysis shows
that it is much more general. What is even more striking is his conclusion about the \emph{difference} in these two
specific heats. By the reasoning given above, this difference must be $X\,\frac{V}{267}$. While the number $\frac{1}{116}$
was for air only, the number $\frac{1}{267}$, by Gay-Lussac law, is the \emph{same} for all ideal gases. Thus, the difference
in the specific heats is completely \emph{insensitive} to the details of the individual gases. In fact, a little
introspection shows that Carnot need not have used Poisson's data at all!

In the next step of the reasoning too, Carnot displays absolute brilliance. He considers two ideal heat engines working
with \emph{different volumes} and just by using some properties of the ideal gas equation such as that for a given fractional
change of pressure at the same temperature produces the same fractional change of volume etc., he demonstrates that $X\cdot V$
is the \emph{same function of temperature} for all gases. Thus, the difference in the specific heat at constant pressure
and the specific heat at constant volume being equal to $X\,\frac{V}{267}$, \emph{it is the same for all ideal gases} at
a given temperature, and is independent of the density.

Introduction of symbolic manipulation already makes the above arguments, though correct, more \emph{transparent}. Let us,
for ease of presentation, consider \emph{one mole} of a gas. The heating at constant pressure, leading to an increase of
temperature by $1 \celsius$, and the adiabatic compression for air also leading to an elevation of the temperature by
$1 \celsius$ can be described by the simple equations
\begin{equation}
\label{eq:carnotcpadia}
Q(P,V+\frac{V}{267},1)-Q(P,V,0) = C_P\quad\quad Q(P^\prime, V-\frac{V}{116},1)=Q(P,V,0)
\end{equation} 
The heating by one degree at constant volume is likewise described by
\begin{equation}
\label{eq:carnotcv}
Q(P^{\prime\prime},V,1)-Q(P,V,0) = C_V
\end{equation}
It immediately follows that
\begin{eqnarray}
\label{eq:carnotcdiff}
C_V &=& Q(P^{\prime\prime},V,1)-Q(P^\prime,V\frac{115}{116},1)\approx \tpd{Q}{V}{T}{}V\frac{1}{116}\nonumber\\
C_P&=& Q(P,V\frac{268}{267},1)-Q(P^\prime,V\frac{115}{116},1)\approx \tpd{Q}{V}{T}{}V(\frac{1}{116}+\frac{1}{267})
\end{eqnarray}
We have symbolized Carnot's principal axiom that \emph{heat is a state function} by using $Q(P,V,T)$. The factor X introduced
earlier is precisely $\tpd{Q}{V}{T}{}$.

\section{Infinitesimal and Finite Cycles} 
\label{sec:composecarnot}
In the above, changes of volumes and temperatures were very small. Let us now discuss Carnot's novel, and extremely useful, concept of
\emph{infintesimal reversible cycles}. These are reversible cycles where each of the four stages is infinitesimally small. He argues that any 
\emph{finite reversible cycle} can be shown to be \emph{equivalent} to a large number of suitably chosen infinitesimal cycles.
\begin{figure}
\begin{center}
\includegraphics[width=3.5in]{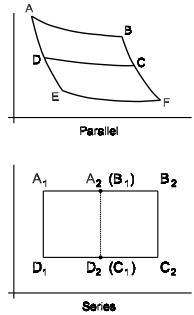}
\end{center}
\caption{Composing Carnot cycles}
\label{fig:carnotadd}
\end{figure}
We illustrate how two Carnot cycles operating between the same two temperatures $T_H, T_L$ can be combined into a single
Carnot cycle. Consider two such cycles $A_1B_1C_1D_1A_1$ and $A_2B_2C_2D_2A_2$ as shown in the next figure, such that
the state $A_2$ is the same as the state $B_1$, and $D_2$ the same as $C_1$. 
We can represent each cycles by the \emph{ordered} set of its segments; for example, $A_1B_1C_1D_1A_1$ can be represented by $A_1B_1,B_1C_1,C_1D_1,D_1A_1$. A segment $B_1C_1$ is to be understood as the \emph{thermodynamic reverse} of the path $A_1B_1$. 
Now the two cycles can be composed by
considering the sequence of paths $A_1B_1, B_1C_1, C_1D_1, D_1A_1)+A_1B_1(A_2)+(A_2B_2,B_2C_2,C_2D_2,D_2A_2)+A_2(B_1)A_1$.
On the one hand, this is the sum of the two given Carnot cycles on recognizing that $A_1B_1(A_2)$ \emph{cancels} $A_2(B_1)A_1$
due to perfect reversibility of the ideal cycles. On the other hand, in the total path, the segment $B_1C_1$ cancels $D_2A_2$,
leading to its reorganization as $A_1A_2,A_2B_2,B_2C_2,C_2D_2(C_1),C_1D_1, D_1A_1$ which is the \emph{composite cycle}. This
can be repeated for composing cycles operating between different combinations of temperatures. It is very important to notice
that reversibility is the key to this composition of cycles.

The abovementioned way of composing Carnot cycles can be, by borrowinig an obvious analogy from electrostaics, called 
a composition in \emph{series}, and is shown as the bottom part in figure~(\ref{fig:carnotadd}). However, one
can also introduce the notion of composing Carnot cycles in \emph{parallel}. In such an arrangement, the lower temperature
of the first cycle would be the same as the higher temperature of the second cycle etc.. This is shown as the top part
of the same figure.

\section{Gas Engines} 
Now we present the analysis of Carnot's \emph{gas engines}, not in his original prosaic style, but in the
succinct mathematical form used by Kelvin
. As explained above, it suffices to analyze an infinitesimal cycle. Let (P,V,T) be
the initial state and let dQ be the 
heat absorbed during the first isothermal stage, and let dV be the corresponding increase in volume, so that the state B
at the end of the first stage is (P(1-$\frac{dV}{V}$),V(1+$\frac{dV}{V}$),T). The mean pressure during stage (i) is therefore
$P(1-\frac{dV}{2V})$ and the work done during this stage is $dV\cdot P(1-\frac{dV}{2V})$. We need to calculate to second order
in accuracy.

During the second stage,
let $\delta P, \delta V and \delta T$ be the decrease in pressure, increase in volume and decrease in temperature, respectively.
Hence the state C is $(P(1-\frac{dV}{V})-\delta P, V(1+\frac{dv}{V})+\delta V, T-\delta T)$. It is a good approximation, as can be
checked easily, to treat the corresponding variations during (iii) and (iv) to be the same as during (i) and (ii).
The ideal gas law, for one mole of gas, then requires
\begin{equation}
\label{eq:kelvincarnot}
-V\,\delta P+P\,\delta V = -R\,\delta T
\end{equation}
In fact, adiabaticity further restricts these variations, but as Kelvin
has rightly remarked, it is not necessary to know them.
The mean pressure during (iii) is therefore $\delta P$ less than the mean pressure during (i), and the net work done during the
isothermal stages is simply $dV\delta P$. The mean pressure during (ii) is therefore $P(1-\frac{dV}{V})-\frac{\delta P}{2}$.
The mean pressure during (iv) is likewise $P\frac{dV}{V}$ \emph{more} than that during (ii(, and the net work done during
the adiabatic stages is $-P\frac{dV}{V}\delta V$. The \emph{total} work done during the cycle is, theefore, $(V\delta P-P\delta v)\frac{dV}{V}$.
On using eqn.(\ref{eq:kelvincarnot}), this can be simplified as $dW = \frac{R}{V}\delta T\,dV$. Following Kelvin
, this is further
reexpressed as
\begin{equation}
\label{eq:kelvincarnotmu}
dW  \equiv \mu(T)\,dQ\delta T = \frac{R}{V\tpd{Q}{V}{T}{}}\,dQ\delta T 
\end{equation}
This is the result that Carnot sought to find, and it expresses the \emph{motive power} dW that the \emph{thermal agency} $dQ\delta T$
will give rise to. According to the powerful \emph{universality} argument of Carnot, the function $\mu(T)$ is the \emph{same} universal
function no matter how the heat engine is designed, or with what substance. For ideal gases, the above mentioned derivation yields
$\mu(T)=\frac{R}{V\tpd{Q}{V}{T}{}}$.

\section{Steam Engines and the Clapeyron Equation} 
\label{sec:clapeyron}
As already mentioned before, Carnot seems to have been under the impression that for 
steam engines, a fully reversible cycle can not be maintained. He based this on the premise that after the steam has condensed to 
water at the lower operating temperature, the water would have to be heated to be at the starting point of the cycle. It was Clapeyron,
in 1834, two years after the untimely death of Carnot(he died in a Cholera epidemic at the age of 36), that showed that
the ideal steam engine can also be thought of as a reversible cycle with the same four stages that Carnot had
given for the gas engine, provided important features of liquid-vapor equilibrium are
taken into account. One of these is that in the P-V diagram for steam engines, the isotherms are at constant pressure because
saturated vapor pressure depends only on temperature. The other is that water can absorb heat to become steam \emph{without}
any change of temperature. The adiabatic curves are basically the same as the P-T
diagrams of coexistence. We now present Clapeyron's analysis of the motive power of steam engines. In this work, Clapeyron
puts to use, in an eloquent way, his \emph{graphical method}, which we have already discussed.

Again, let us consider only an infinitesimal cycle EFGH shown as a horizontal strip in the figure. The work done is given by the
area of this strip which is, to a good approximation, the length EF multiplied by dP which is the thickness. The length EF is
essentially the change in total volume of the system upon absorbing the amount of heat dQ. If $l(T)$ is the \emph{latent heat}
(in the modern sense i.e amount of heat required to convert unit mass of water at temperature T to unit mass of steam at the 
same temperature; in Carnot's times the phrase \emph{latent heat} was used in a different sense), the mass dm of water converted
to steam is $dm =\frac{dQ}{l(T)}$. The increase in volume of steam is therefore $dV_{steam} = \frac{dm}{\rho_s}$, where
$\rho_s$ is the density of steam. No heat is lost to the water as neither its pressure nor temperature changes. However, there
is mass loss of water, also by dm. This leads to a decrease in the volume of water by $dV_{water} = -\frac{dm}{\rho_w}$, where
$\rho_w$ is now the density of water. Both the densities depend on T. Therefore, $EF = dV =dm(v_s-v_w)$, where $v_s,v_w$
are the \emph{specific volumes} i.e volume per unit mass of steam and water, respectively. Consequently, the work done during the cycle is 
$dW_s =EF\cdot dP=\frac{1}{l(T)}\,(v_s-v_w)\,dPdQ$. This can be rewritten as follows:
\begin{equation}
\label{eq:kelvinmusteam}
dW_s = \left\{\frac{v_s-v_w}{l(T)}\,\frac{dP(T)}{dT}\right\}dQdT\quad\rightarrow \mu(T) = \left\{\frac{v_s-v_w}{l(T)}\,\frac{dP(T)}{dT}\right\}
\end{equation}
Now one can appreciate the true powers of the universality of ideal heat engines propounded by Carnot. According to it, 
$\mu(T)$ is the \emph{same} function of temperature for \emph{all} substances. The implication for steam-water coexistence can
be deduced by rewriting the above equation as
\begin{equation}
\label{eq:clapeyronorig}
\frac{dP(T)}{dT} = \mu(T)\,\frac{l(T)}{v_s-v_w}
\end{equation}
This is the famous \emph{Clapeyron Equation} and it has been obtained from the Caloric theory! The missing ingredient, however,
was the function $\mu(T)$, and even Clapeyron bemoans the lack of reliable experimental data that would determine it. Regnault's
careful work on steam \cite{regnault:1847}, which Kelvin
made use of at the time of his commentary on Carnot's work, would only start to
become available in 1847, the full descriptions completed as late as 1870.

Returning to the specific heats of
ideal gases, one gets
\begin{equation}
\label{eq:carnotcpcv}
(C_P-C_V)(0) = \frac{R}{267\mu(0)}
\end{equation}
We shall now go a step beyond Kelvin
, and give a completely mathematical treatment of Carnot's work.

\section{Mathematical treatment of Carnot theory} 
The starting point of Carnot's considerations was the \emph{Caloric Theory},
which states that heat is a property of the system. More precisely, it states that \emph{heat is a state function}, and mathematically
this amounts to the existence of the \emph{heat function} $Q(V,T)$. It can equally well be expressed as $Q(P,V)$ or $Q(P,T)$.
As we have already seen, for ideal gases $V\tpd{Q}{V}{T}{} = \frac{R}{\mu(T)}$. The \emph{Holy Grail} of Carnot theory is
the determination of both $Q(V,T)$ and $\mu(T)$. Of course, knowing Q(V,T) for ideal gases at once gives $\mu(T)$ which holds for
\emph{all} substances. We develop the mathematical theory for ideal gases here, but it can be extended to arbitrary cases. 

Let us consider specific heat at constant volume $C_V$(we consider one mole of the substance). By definition, the heat dQ required
to raise the temperature by dT is $C_V(V,T)dT$. We leave open the possibility that the specific heats could depend on (V,T). In the
caloric theory
\begin{equation}
\label{eq:mathcv}
C_VdT = Q(P^{\prime\prime}, V, T+dT)-Q(P,V,T) = \tpd{Q}{T}{V}{}dT \rightarrow C_V=\tpd{Q}{T}{V}{}
\end{equation}
But Carnot finds it more useful to understand $C_V$ in terms of heat required to change volumes \emph{at constant temperature}!
That he does by invoking the properties under \emph{adiabatic} changes. Let $\delta_{ad} V$ be the change in volume, under
adiabatic changes, corresponding to a change $\delta_{ad} T$ in temperature. For air, considered by Carnot for which he quotes
the experiments of Poisson, $\delta_{ad} V = -\frac{V}{116}$ when $\delta_{ad} T = 1\celsius$. The mathematical expression
for adiabatic changes in the caloric theory is
\begin{equation}
\label{eq:mathadia}
Q(P,V,T)-Q(P^\prime, V+\delta_{ad}V,T+\delta_{ad}T)=0\rightarrow \tpd{Q}{V}{T}{}\delta_{ad}V+\tpd{Q}{T}{V}{}\delta_{ad}T=0
\end{equation}
This is the same conclusion reached by Carnot, namely, \emph{the heat absorbed at constant temperature in expanding by a small
volume is the same as would be required to raise the temperature, at constant volume, by a degree by which the temperature
would have increased under adiabatic compression by the same volume}. What is noteworthy is that Carnot arrives at it through
only verbal manipulations!

The second important assertion by him, again proved only verbally, is that \emph{the heat given out, at constant temperature,
only depends on the fractional increase in volume and not on the increase in volume itself}. To arrive at that conclusion,
he makes use of his result on the universality of ideal heat engines. In the mathematical formalism this emerges as
follows:
\begin{equation}
\label{eq:carnotsecond}
dq=\tpd{Q}{V}{T}{}\,dV = \frac{R}{\mu(T)}\frac{dV}{V}
\end{equation}
In fact, Carnot enunciates this result for \emph{finite} changes as well(also proved verbally!): \emph{'When a gas varies in volume
without change of temperature, the quantities of heat absorbed or liberated by this gas are in arithmetical progression, if the increments
or decrements in volume are found to be in geometrical progression}. To see this in our mathematical formulation, simply integrate
eqn.(\ref{eq:carnotsecond}), to give,
\begin{equation}
\label{eq:carnotfinite}
Q_2-Q_1 = \frac{R}{\mu(T)}\,\ln \frac{V_2}{V_1}
\end{equation} 
Though Carnot used Poisson's data for air on adiabatic changes, he could well have made that analysis more general as the law
for adiabatic changes, in the form, $PV^\gamma = const.$, appears to have been known to Laplace, whose work on speed of sound
is cited by Carnot. But, as can be seen now, the mathematical theory of the caloric gives the equivalent of this relation even
when the specific heats are not constant.

To address this and other related issues , let us turn our attention to the specific heats within the caloric theory. One of the differential 
forms of the fundamental axiom of the caloric theory can be expressed as:
\begin{equation}
\label{eq:caloricdiff}
dQ = \tpd{Q}{P}{V}{}dP+\tpd{Q}{V}{P}{}dV
\end{equation}
Other equivalent forms using (P,T) or (V,T) as independent variables may also be used. From the definitions $C_V=\tpd{Q}{T}{P}{}$
and $C_V=\tpd{Q}{T}{V}{}$, it immediately follows that for ideal gases
\begin{equation}
\label{eq:mathcpcv}
C_P =\tpd{Q}{V}{P}{}\tpd{V}{T}{P}{}=\frac{R}{P}\,\tpd{Q}{V}{P}{}\quad\quad C_V =\frac{R}{V}\,\tpd{Q}{P}{V}{}
\end{equation}
The ratio, $\gamma$, of $C_P$ to $C_V$ in the caloric theory is given by
\begin{equation}
\label{eq:caloricgamma}
\gamma(V,T) = \frac{C_P(V,T)}{C_V(V,T)}= \frac{V}{P}\frac{\tpd{Q}{V}{P}{}}{\tpd{Q}{P}{V}{}}=\frac{V}{P}\tpd{Q}{V}{P}{}\tpd{P}{Q}{V}{}
\end{equation}
Using the triple product rule of partial derivatives, one obtains
\begin{equation}
\tpd{Q}{V}{P}{}\tpd{P}{Q}{V}{}=-\tpd{P}{V}{Q}{}\rightarrow \frac{\delta_{ad}P}{P}+\gamma(V,T)\,\frac{\delta_{ad}V}{V}=0
\end{equation}
Which is incidentally the same equation for adiabatic changes in modern thermodynamics too. Therefore, this particular equation
does not care what the nature of heat is.

Let us evaluate $\tpd{Q}{V}{T}{}$ for an ideal gas directly from eqn.(\ref{eq:caloricdiff}):
\begin{equation}
\label{eq:mathcdiff}
\tpd{Q}{V}{T}{} = \tpd{Q}{V}{P}{}+\tpd{Q}{P}{V}{}\tpd{P}{V}{T}{}=\frac{P}{R}C_P-\frac{P}{R}C_V
\end{equation}
Combining this with the expression for $\mu(T)$, one gets the remarkable equality
\begin{equation}
\label{eq:mathcdiff2}
C_P-C_V = \frac{R}{\mu(T)T}
\end{equation}
This is the mathematical derivation of Carnot's result for the specific heats; 
and the difference can only depend on temperature, with $C_P$ always \emph{greater} than $C_V$. Carnot had
concluded that if $C_P-C_V$ was a \emph{constant}, the specific heats must have a \emph{logarthmic} dependence on
volume. In our mathematical framework, this case amounts to fixing $\mu(T)$ to be $\frac{1}{T}$ i.e hence 
$\tpd{Q}{V}{T}{}=\frac{RT}{V}$, whose solution is $Q(V,T) = RT\ln V +f(T)$ with $f(T)$ being arbitrary.
Therefore, $C_V(V,T) = R\ln V +f^\prime(T)$ and $C_P(V,T) = R\ln V+f^\prime(T)+R$.
One can likewise explore the consequences of a \emph{constant} $C_V$. 
It is easy to see that this would imply $Q(V,T) = C_V\,T+f(V)$,
$f(V)$ being arbitrary. Then, $C_P=C_V+f^\prime(V)\,\frac{R}{P}$. But $C_P-C_V$
can only be a function of T which fixes $f^\prime(V) = \frac{A}{V}$ with A a constant. Consequently $C_P=C_V+\frac{A}{T}$.

Finally, we present the differential form of the caloric axiom for ideal gases in a form that is closest to the present
day first law. For that, we take (V,T) as the independent variables:
\begin{equation}
\label{eq:caloricfirst}
dQ(V,T) = \tpd{Q}{T}{V}{}dT+\tpd{Q}{V}{T}{}dV = C_V(V,T)dT+\frac{R}{\mu(T)}\frac{dV}{V}
\end{equation}

Carnot was very particular in his views about the importance of subjecting his conclusions to rigorous experimental tests. He
correctly foresaw specific heat data to be the most important ones for this purpose. But the state of the art of these experiments
were not fine enough, and in fact, the data of Clement and Desormes which made Carnot see some evidence for a logarthmic volume
dependence were later found to be incorrect. It is undoubtedly clear that had Carnot lived to see greater precision in these
experiments, he would have been the first to abandon the caloric theory, and perhaps the first to have formulated the first and
second laws of thermodynamics! After all, the important ideas of Carnot and Clapeyron, in the hands of Clausius, paved the
way for these developments. However, despite his great contributions, particularly the concepts of reversible cycles, universality
of efficiencies, and of maximum of attainable efficiencies, it can not be said that he knew of even the broad contours of the first 
and second laws as understood today. The reader is referred to \cite{jsrinivasan:2001}
for a different view. 


\acknowledgments
The author wishes to acknowledge support from the Department of Science and Technology to the project IR/S2/PU-001/2008.

\end{document}